\documentclass[12pt]{article}
\title{A generalisation of the Heckmann - Schucking cosmological solution}
\author{I.M. Khalatnikov$^{1,2,3}$ and A.Yu. Kamenshchik$^{1,2}$ }
\date{}
\begin{document}
\maketitle
\hspace{-5mm}$^1$L.D. Landau Institute for Theoretical Physics of Russian
Academy of Sciences, Kosygin str. 2, 117334, Moscow, Russia\\
$^2$Landau Network - Centro Volta, Villa Olmo, via Cantoni 1, 22100 Como,
Italy\\
$^{3}$Tel Aviv University,
Tel Aviv University, Raymond and Sackler
Faculty of Exact Sciences, School of Physics and Astronomy,
Ramat Aviv, 69978, Israel\\

\begin{abstract}
An exact solution of the Einstein equations for a Bianchi -I universe 
in the presence of dust, stiff matter and cosmological constant, generalising the well-known Heckmann-Schucking solution is presented.
\end{abstract}
PACS: 04.20-q; 04.20.Dw\\
Keywords: Exact cosmological solutions
\section{Introduction}
In spite of the development of the numerical methods for integration 
of the Einstein equations and an extensive application of the qualitative 
theory of differential equations to analysis of the behavior of the cosmological models, the construction of exact solutions still represents 
rather an attractive task. One can underline two important classes 
of the cosmological solutions known now. One of them is the class of exact solutions of Friedmann-Robertson-Walker isotropic cosmological 
models, which constitute a basis for comparison of theoretical 
predictions with observations \cite{Land-Lif}. Another important exact 
cosmological solution is the anisotropic Kasner solution for the empty 
Bianchi-I universe \cite{Kasner}, whose importance for theoretical 
physics is connected with its role in the description of the oscillatory 
aapproach to the cosmological singularity \cite{BKL}, which in turn 
appears very promising topic for the study in the string and M-theory 
context \cite{string}.

The Heckmann - Schucking \cite{Schuck} anisotropic solution for the  Bianchi-I universe in the presence of the dust-like matter is of special interest because it constitutes some kind of bridge between these two types of the cosmological solutions: in the vicinity of the cosmological 
singularity it behaves as a Kasner universe, while at the later stage of
the cosmological evolution it behaves as an isotropic flat Friedmann 
Universe. The Heckmann-Schucking solution has the following form:
for the Bianchi-I universe with the metric  
\begin{equation}
ds^2 = dt^2 - a^2(t)dx^2 - b^2(t)dy^2 - c^2(t)dz^2
\label{metric}
\end{equation}
filled with dust whose equation of state is
\begin{equation}
p = 0
\label{dust}
\end{equation}
the functions $a(t), b(t)$ and $c(t)$ are given by the formulae
\begin{eqnarray}
&&a(t) = a_0t^{p_1}(t+t_0)^{2/3-p_1},\nonumber \\
&&b(t) = b_0t^{p_2}(t+t_0)^{2/3-p_2},\nonumber \\
&&c(t) = c_0t^{p_3}(t+t_0)^{2/3-p_3},
\label{Schuck}
\end{eqnarray}
where the exponents $p_1, p_2$ and $p_3$ are the well-known Kasner exponents \cite{Land-Lif,BKL}
satisfying relations
\begin{equation}
p_1 + p_2 + p_3 = 1,
\label{Kasner}
\end{equation}
\begin{equation}
p_1^2 + p_2^2 + p_3^2 = 1.
\label{Kasner1}
\end{equation}
It is easy to see that the solution (\ref{Schuck}) is close to the Kasner solution when $t \ll t_0$. In the limit $t \gg t_0$ all the functions 
$a(t), b(t)$ and  $c(t)$ are proportional to $t^{2/3}$, i.e. their behavior 
 coincides with that of the flat Friedmann universe filled with the dust.
The energy density of the matter is given by the formula
\begin{equation}
\varepsilon = \frac{E_0}{t(t+t_0)},
\label{dust1}
\end{equation}
where the constant $E_0 \neq 0$ and does not depend on the choice of 
the Kasner exponents. 

In this paper we generalise the Heckmann - Schucking solution for the case of the Bianchi - I universe filled with the mixture of three perfect fluids: dust, stiff matter and a cosmological constant. The presence of the cosmological constant in combination with other types of matter is of special interest, because the recent discovery of the phenomenon of the cosmic acceleration \cite{cosmic} makes the presence of the cosmological 
constant or some other exotic type of matter which mimics some basic 
features of the cosmological constant enavoidable for any realistic cosmological model.  Another reason which makes the study of the models with the cosmological constant interesting is the fact that 
inflationary theories of a very early Universe \cite{inflation} contain 
an effective cosmological constant providing a period of a quasi-exponential expansion at the beginning of the cosmological evolution.

The plan of the paper is the following one: in second section we integrate 
the Einsten equations for the Bianchi - I model filled with the mixture of three perfect fluids mentioned above; in the third section we consider different limit cases for our solution and rewrite it in the form maximally close to the canonical form of the Heckmann-Schucking solution (\ref{Schuck}).

\section{Integration of the Einstein equations for a Bianchi - I cosmology in the presence of matter}
We shall look for solutions of the Einstein equations for the Bianchi - I universe with the metric (\ref{metric}) filled with the matter with an isotropic matter with the energy-momentum tensor which has the form
\begin{equation}
T_{\mu}^{\nu} = diag(\varepsilon, -p, -p, -p).
\label{EMT}
\end{equation}
It is convenient to represent the fucntions $a(t), b(t)$ and $c(t)$ in the following form:
\begin{eqnarray}
&&a(t)  = R(t)\exp(\alpha(t) + \beta(t)), \nonumber \\
&&b(t) = R(t)\exp(\alpha(t) - \beta(t)), \nonumber \\
&&c(t) = R(t)\exp(-2\alpha(t)),
\label{assym}
\end{eqnarray}
where $R(t)$ is the conformal factor while the functions $\alpha(t)$ and $\beta(t)$ characterise the anisotropy of the model. The components of the Ricci tensor have the following form:
\begin{equation}
R_0^0 =-\left(3 \frac{\ddot{R}}{R} + 6 \dot{\alpha}^2 + 2\dot{\beta}^2\right),
\label{Ricci0}
\end{equation}
\begin{equation}
R_1^1 =-\left( \frac{\ddot{R}}{R} + 2\frac{\dot{R}^2}{R^2} + 3\frac{\dot{R}}{R}(\dot{\alpha} + \dot{\beta}) 
+ \ddot{\alpha} + \ddot{\beta}\right),
\label{Ricci1}
\end{equation}
\begin{equation}
R_2^2 =-\left( \frac{\ddot{R}}{R} + 2\frac{\dot{R}^2}{R^2} + 3\frac{\dot{R}}{R}(\dot{\alpha} - \dot{\beta}) 
+ \ddot{\alpha} - \ddot{\beta}\right),
\label{Ricci2}
\end{equation}
\begin{equation}
R_3^3 =-\left( \frac{\ddot{R}}{R} + 2\frac{\dot{R}^2}{R^2} -6\frac{\dot{R}}{R}\dot{\alpha} 
- 2\ddot{\alpha} \right).
\label{Ricci3}
\end{equation}
Due to the isotropy of the energy-momentum tensor (\ref{EMT}) one has
\begin{equation}
R_1^1 = R_2^2 = R_3^3.
\label{Einstein}
\end{equation} 
From this equation it is easy to get equations describing for the asymmetry functions $\alpha(t)$ and $\beta(t)$:
\begin{equation}
\ddot{\alpha} + 3\frac{\dot{R}}{R}\dot{\alpha} = 0,
\label{alpha}
\end{equation}
\begin{equation}
\ddot{\beta} + 3\frac{\dot{R}}{R}\dot{\beta} = 0.
\label{beta}
\end{equation}
The form of Eqs. (\ref{alpha}) and (\ref{beta}) coincides with that for the Klein - Gordon equation for the massless sclalar field on a Friedmann background, whose energy-momentum tensor coincides with the energy-momentum tensor of the stiff matter \cite{BK} with the equation of state
\begin{equation}
p = \varepsilon.
\label{stiff} 
\end{equation}
From Eqs. (\ref{alpha}), (\ref{beta}) it follows immediately that
\begin{equation}
\dot{\alpha} = \frac{\alpha_0}{R^3},
\label{alpha1}
\end{equation}
\begin{equation}
\dot{\beta} = \frac{\beta_0}{R^3},
\label{beta1}
\end{equation}
where $\alpha_0$ and $\beta_0$ are some positive constants. 

The $00$ component of the Einstein equations has now the form
\begin{equation}
\frac{\dot{R}^2}{R^2} = \dot{\alpha}^2 + \frac{\dot{\beta}^2}{3} 
+ \frac{4\pi G}{3}\varepsilon.
\label{Einstein0}
\end{equation}
We shall consider a universe filled with  a mixture of three perfect fluids:
dust, the stiff matter obeying the equation of state (\ref{stiff}) and the cosmological constant, whose equation of state is $p = -\varepsilon$. 
Choosing convenient normalisation of constants one can represent 
Eq. (\ref{Einstein0}) for the universe filled with this mixture in the following form:
\begin{equation}
\frac{\dot{R}^2}{R^2} = \dot{\alpha}^2 + \frac{\dot{\beta}^2}{3}
+ \Lambda + \frac{M}{R^3} + \frac{S}{R^6},
\label{Friedmann}
\end{equation}
where $\Lambda$ is the cosmological term, while the constants $M$ and $S$ characterise the quantity of the dust and of the stiff matter in the universe respectively.

After substitution into Eq. (\ref{Friedmann}) the expressions for $\dot{\alpha}$ and $\dot{\beta}$ from Eqs. (\ref{alpha1}) and (\ref{beta1}) we come to the following equation for the conformal factor $R(t)$
\begin{equation}
\frac{\dot{R}^2}{R^2} = 
\Lambda + \frac{M}{R^3} + \frac{S_0}{R^6},
\label{Friedmann1}
\end{equation}
 where 
\begin{equation}
S_0 = S + \alpha_0^2 + \frac{\beta_0^2}{3}.
\label{tildeS}
\end{equation}

It is possible to integrate Eq. (\ref{Friedmann1}) explicitly and the result 
is the following one:
\begin{equation}
R^3(t) = \frac{M}{2\Lambda}(\cosh 3\sqrt{\Lambda}t - 1) + \sqrt{\frac{S_0}{\Lambda}}\sinh 3\sqrt{\Lambda}t.
\label{Friedmann2}
\end{equation}
The initial conditions for the solution (\ref{Friedmann2}) are chosen in such a way to provide the fulfilling of the relation $R(0) = 0$.
Substituting the solution (\ref{Friedmann2}) into Eqs. (\ref{alpha1}) and 
(\ref{beta1}) one has after taking integrals:
\begin{equation}
\alpha(t) = \frac{\alpha_0}{3\sqrt{S_0}}\ln \left(\frac{e^{3\sqrt{\Lambda}t} - 1}{e^{3\sqrt{\Lambda}t} + \frac{2\sqrt{S_0\Lambda}-M}{2\sqrt{S_0\Lambda}+M}}\right),
\label{alpha2}
\end{equation}
\begin{equation}
\beta(t) = \frac{\beta_0}{3\sqrt{S_0}}\ln \left(\frac{e^{3\sqrt{\Lambda}t} - 1}{e^{3\sqrt{\Lambda}t} + \frac{2\sqrt{S_0\Lambda}-M}{2\sqrt{S_0\Lambda}+M}}\right).
\label{beta2}
\end{equation}
In the expressions (\ref{alpha2}), (\ref{beta2}) we have omitted the integration constants, inclusion of which implies only the multiplication 
of factors $a(t), b(t)$ and $c(t)$ by some factors, that is not important for the flat (Bianchi - I) model.     

Now, substituting the formulae (\ref{Friedmann2}), (\ref{alpha2}) and (\ref{beta2}) into Eq. (\ref{assym}) one comes to the following cosmological solution:
\begin{eqnarray}
&&a(t) = \left(\frac{M}{\Lambda}\right)^{\frac13}\left(1 + \frac{2\sqrt{S_0\Lambda}}{M}\right)^{\frac{\alpha_0+\beta_0}
{3\sqrt{S_0}}}\left(\sinh\frac{3\sqrt{\Lambda}t}{2}\right)^
{\frac13 + \frac{\alpha_0+\beta_0}{3\sqrt{S_0}}}\nonumber \\
&&\times \left(\sinh\frac{3\sqrt{\Lambda}t}{2} + 
\frac{2\sqrt{S_0\Lambda}}{M}\cosh\frac{3\sqrt{\Lambda}t}{2}
\right)^ {\frac13 - \frac{\alpha_0+\beta_0}{3\sqrt{S_0}}}, \nonumber \\
&&b(t) = \left(\frac{M}{\Lambda}\right)^{\frac13}\left(1 + \frac{2\sqrt{S_0\Lambda}}{M}\right)^{\frac{\alpha_0-\beta_0}
{3\sqrt{S_0}}}\left(\sinh\frac{3\sqrt{\Lambda}t}{2}\right)^
{\frac13 + \frac{\alpha_0-\beta_0}{3\sqrt{S_0}}}\nonumber \\
&&\times \left(\sinh\frac{3\sqrt{\Lambda}t}{2} + 
\frac{2\sqrt{S_0\Lambda}}{M}\cosh\frac{3\sqrt{\Lambda}t}{2}
\right)^ {\frac13 - \frac{\alpha_0-\beta_0}{3\sqrt{S_0}}}, \nonumber \\
&&c(t) = \left(\frac{M}{\Lambda}\right)^{\frac13}\left(1 + \frac{2\sqrt{S_0\Lambda}}{M}\right)^{\frac{-2\alpha_0}
{3\sqrt{S_0}}}\left(\sinh\frac{3\sqrt{\Lambda}t}{2}\right)^
{\frac13 + \frac{-2\alpha_0}{3\sqrt{S_0}}}\nonumber \\
&&\times \left(\sinh\frac{3\sqrt{\Lambda}t}{2} + 
\frac{2\sqrt{S_0\Lambda}}{M}\cosh\frac{3\sqrt{\Lambda}t}{2}
\right)^ {\frac13 + \frac{2\alpha_0}{3\sqrt{S_0}}}.
\label{solution}
\end{eqnarray}
The solution  presented above has rather a cumbersome form. In the next section we consider different limit cases and rewrite Eq. (\ref{solution}) in a more convenient Heckmann-Schucking-like form. 

\section{Limit cases and the Heckmann-Schucking-like form of the anisotropic cosmological solution}
Let us consider the limit case $\Lambda = 0$ in Eq. (\ref{solution}), i.e. the case when the cosmological constant in absent. One has the following expressions:
\begin{eqnarray}
&&a(t) = \left(\frac94 M\right)^{\frac13} t^{p_1}
(t + t_0)^{\frac23 - p_1}, \nonumber \\
&&b(t) = \left(\frac94 M\right)^{\frac13} t^{p_2}
(t + t_0) ^{\frac23 - p_2}, \nonumber \\
&&c(t) = \left(\frac94 M\right)^{\frac13} t^{p_3}
(t + t_0) ^{\frac23 - p_3}.
\label{solution1}
\end{eqnarray}
Here we have introduced the parameters  
\begin{equation}
t_0  = \frac{4\sqrt{S_0}}{3M},
\label{t0}
\end{equation}
\begin{eqnarray}
&&p_1 = \frac13 - \frac{\alpha_0+\beta_0}{3\sqrt{S_0}},\nonumber \\
&&p_2 = \frac13 - \frac{\alpha_0-\beta_0}{3\sqrt{S_0}},\nonumber \\
&&p_3 = \frac13 + \frac{2\alpha_0}{3\sqrt{S_0}}.
\label{Kasnernew}
\end{eqnarray}
The parameter $t_0$ plays the same role as in the 
Heckmann-Schucking solution (\ref{Schuck}), but now it depends not only on the 
anisotropy parameters $\alpha_0$ and $\beta_0$ but also on the quantity of the stiff matter in the universe(see Eq. (\ref{tildeS})). It is easy to see that the parameters $p_1, p_2$ and $p_3$, which  can be called ``quasi-Kasner'' parameters satisfy the relations
\begin{eqnarray}
&&p_1 + p_2 + p_3 = 1, \\
\label{Kasnernew0}
&&p_1^2 + p_2^2 + p_3^2 = 1 - q^2, 
\label{Kasnernew1}
\end{eqnarray}
where
\begin{equation}
q^2 = \frac23 \frac{S}{\alpha_0^2 + \frac{\beta_0^2}{3}+S} = \frac{2S}{3S_0}.
\label{equation}
\end{equation}
The parameter $q$ reflects the presence of the stiff matter in a universe
and was introduced in \cite{BK}.  
Apparently, when the stiff matter is absent $q = 0$ and the solution (\ref{solution1}) exactly coincides with the Heckmann-Schucking solution (\ref{Schuck}) for a Bianchi - I universe filled with dust. 
The presence of the stiff matter simply transforms the relation  
(\ref{Kasner1}) for the Kasner exponents into Eq. (\ref{Kasnernew1}).  

Now, it convenient to rewrite the solution (\ref{solution}) using the Heckmann - Schucking time parameter $t_0$ and the quasi-Kasner
exponents. We shall use also the parameter 
\begin{equation}
H_0 = \sqrt{\Lambda},
\label{Hubble}
\end{equation}
which is nothing but a Hubble parameter for the DeSitter universe with a 
cosmological constant $\Lambda$.
Now the solution (\ref{solution}) looks like
\begin{eqnarray}
a(t) = \left(\frac{M}{H_0^2}\right)^{\frac13}\left(1+\frac{3 H_0 t_0}{2}\right)^{p_1 - \frac13}\left(\sinh \frac{3 H_0 t}{2}\right)^{p_1}
\left(\sinh \frac{3 H_0 t}{2} + \frac{3 H_0 t_0}{2} \cosh \frac{3 H_0 t}{2}\right)^{\frac23 - p_1},&& \nonumber \\
b(t) = \left(\frac{M}{H_0^2}\right)^{\frac13}\left(1+\frac{3 H_0 t_0}{2}\right)^{p_2 - \frac13}\left(\sinh \frac{3 H_0 t}{2}\right)^{p_2}
\left(\sinh \frac{3 H_0 t}{2} + \frac{3 H_0 t_0}{2} \cosh \frac{3 H_0 t}{2}\right)^{\frac23 - p_2},&& \nonumber \\
c(t) = \left(\frac{M}{H_0^2}\right)^{\frac13}\left(1+\frac{3 H_0 t_0}{2}\right)^{p_3 - \frac13}\left(\sinh \frac{3 H_0 t}{2}\right)^{p_3}
\left(\sinh \frac{3 H_0 t}{2} + \frac{3 H_0 t_0}{2} \cosh \frac{3 H_0 t}{2}\right)^{\frac23 - p_3}.&& 
\label{solution2}
\end{eqnarray}

The solution for the Bianchi - I cosmology in the presence of the cosmological constant, dust and stiff matter, presented in the Heckmann-Schucking-like form (\ref{solution2}) is more convenient for the comparison with other solutions. We have already considered the case $\Lambda = 0$ which reproduces the Heckmann-Schucking solution (\ref{Schuck}) up to change of the relation (\ref{Kasnernew1}) between 
the Kasner exponents, which occurs due to the presence of the stiff matter. For completeness we shall write down also the solution for the case when dust component is absent. Taking the corresponding limit in Eq. (\ref{solution2}) we easily come to 
\begin{eqnarray}
&&a(t) = \sinh (3 H_0 t)^{\frac{1}{3}}\left(\tanh \frac{3 H_0 t}{2}\right)^{p_1 - \frac13},\nonumber \\
&&b(t) = \sinh (3 H_0 t)^{\frac{1}{3}}\left(\tanh \frac{3 H_0 t}{2}\right)^{p_2 - \frac13},\nonumber \\
&&c(t) = \sinh (3 H_0 t)^{\frac{1}{3}}\left(\tanh \frac{3 H_0 t}{2}\right)^{p_3 - \frac13}.
 \label{dustless}
\end{eqnarray}

We shall write down also asymptotics of the solution (\ref{solution2}) 
for large and small  times. At large times $t \rightarrow$ only the cosmological constant is relevant and a complete isotropisation takes place. The corresponding asymptotics look like 
\begin{equation}
a(t) \sim \exp(H_0 t),\ b(t) \sim \exp(H_0 t),\ c(t) \sim \exp(H_0 t).
\label{tlarge}
\end{equation} 
At the beginning of the cosmological evolution when $t \rightarrow 0$ only anisotropy and the presence of stiff matter determine the behavior of functions $a(t), b(t)$ and $c(t)$. Their form is now
\begin{equation}
a(t) \sim t^{p_1},\ b(t) \sim t^{p_2},\ c(t) \sim t^{p_3},
\label{tsmall}
\end{equation}
where the exponents $p_1, p_2, p_3$ satisfy as usual the relations (\ref{Kasnernew0}), (\ref{Kasnernew1}). Let us notice that in the model under consideration there are two time scales: $t_0$ and $H_0^{-1}$. 
One can write down also ``intermediate'' asymptotics. In the case when 
$H_0 t_0 < 1$ the solution for the time scale 
\begin{equation}
t_0 < t < \frac{1}{H_0}
\label{inter1}
\end{equation}
looks like 
\begin{equation}
a(t) \sim t^{\frac23},\ b(t) \sim t^{\frac23},\ c(t) \sim t^{\frac23}
\label{inter2}
\end{equation}
and corresponds to the flat Friedmann universe filled with dust.
For the case when 
$H_0 t_0 > 1$ and 
\begin{equation}
\frac{1}{H_0} < t < t_0
\label{inter3}
\end{equation}
the solution coincides with that from Eq. (\ref{tlarge}), i.e. describes the 
flat DeSitter universe.

In conclusion, notice that recently \cite{Schuck1} exact solutions for Bianchi - I universes filled by stiff matter or in the presence of the cosmological constant were considered with use of the original technique \cite{Schuck} using analogies between Einstein and Newton cosmologies.
The qualitative theory for the Bianchi -I universes filled with different 
types of matter was developed in \cite{qualit}.

This work was supported by RFBR via grants No 02-02-16817 and 00-15-96699 and by International Scientific Program "Relativistic Astrophysics" by the Ministry of Industry, Science and Technology of the Russian Federation. The authors are grateful to V.A. Belinsky for useful discussion.

\end{document}